\documentstyle[prl,aps,twocolumn,float]{revtex}

\include{epsf}

\begin{document}

\twocolumn[\hsize\textwidth\columnwidth\hsize\csname
@twocolumnfalse\endcsname

\draft

\title{Spectroscopic Properties and STM Images of Carbon Nanotubes}

\author{Angel Rubio}
\address{
Departamento de F\'{\i}sica Te\'orica, Universidad de Valladolid,
E-47011 Valladolid, Spain.
}

\maketitle

\begin{abstract}
We present a theoretical study
of the role of the local environment in the
electronic properties of carbon nanotubes: 
isolated single- and multi-wall nanotubes, nanotube-ropes,
tubes supported on gold and cutted to finite length.
Interaction with the substrate or with other tubes does not alter the 
scanning-tunneling-microscopy (STM)  patterns observed for isolated tubes. 
A finite length nanotube shows standing-wave 
patterns that can be completely characterized by a set of four different
three-dimensional shapes. These patterns are understood in 
terms of a simple $\pi$-electron tight-binding (TB) model. 
STM-topographic images
of topological defects ani
(pentagon/heptagon pair) and tube-caps have also been studied. 
In both cases the obtained image depends on the sign of the applied voltage 
and it can be described in terms of the previous catalog of STM-images
(interference between electronic waves scattered by the defect).
We also have computed the electronic 
density of states for isolated tubes with different chiralities and radii
confirming a correlation between the peak-structure in
the DOS and the nanotube diameter, however the metallic plateau in the DOS
also depends on the nanotube chirality.  Furthermore, the
conduction and valence band structures are not fully symmetrical
to one another. This anisotropy shows up in the 
DOS and indicates the limitations of the $\pi$-TB model to describe 
spectroscopic data.
In contrast to STM images, the interaction with
the substrate does modify the energy levels of the nanotube.
We observe opening of
small pseudogaps around the Fermi level and broadening of the sharp van Hove
singularities of the isolated single-walled-nanotubes that can be used to
extract useful information about the tube structure and bonding. 
The combination of  STM and spectroscopic studies opens a new technique to 
address the electronic and structural properties of carbon and 
composite nanotubes.
\end{abstract}
]

Carbon is an extraordinary element that appears in a wide variety of 
network-like structures with new potential technological 
applications\cite{book,book1,book2,c_m_n}.
Among these new forms, carbon nanotubes\cite{Iijima} are the most 
promising class of new carbon-based materials for either electronic 
and optic nanodevices as well as composite reinforcement materials. 
The quasi-one dimensional structure and crystallinity of the sample is 
responsible for the unique electronic and mechanical properties of carbon 
nanotubes. The special geometry makes the nanotubes 
excellent candidates for use as nanoscopic quantum wires.

The type of nanotube obtained experimentally
depends on the synthesis conditions and whether a 
catalytic metal particle is used or not.   
Multiwall nanotubes (MWNT) have diameters usually in the range 1-25 nm and
are many microns in length, whereas single-wall nanotube (SWNT)
synthesis requires catalytic particles and SWNTs
have diameters in the range of 0.6-2 nm~\cite{Smalley,Catherine}. 
SWNTs are rarely seen as individual entities and they tend to appear 
in the form of ``ropes'', that usually consist of up to a hundred mono-disperse
nanotubes packed in a perfect triangular lattice. 
Experiments (both dual pulsed-laser~\cite{Smalley} and standard electric
arc technique~\cite{Catherine}) show that the nanotubes forming these 
ropes often have a diameter of approximately 1.3~nm\cite{Smalley,Venema}. 
 
Single-wall carbon nanotubes can de described as single graphene sheet wrapped 
into a seamless cylinder. This structure is completely determined by two
integer numbers $(n,m)$ that defines the circumferential vector 
(${\bf c}=n {\bf a_1} + m {\bf a_2}$) with respect to the two Bravais 
translation vectors of a graphene sheet. The electronic properties of carbon 
nanotubes  are dictated by their geometry  with either semiconducting or 
metallic behavior~\cite{Hamada}. In particular the armchair $(n,n)$ nanotube
with its $n$-mirror planes containing the tubular axis  has
an atomic periodicity along the tubular axis of $a$=2.46~\AA~
and is expected to be metallic with
two $\pi$-bands crossing at the Fermi level.
The relation between nanotube chirality and its electrical properties can be
complementarly explored by theoretical calculations and
Scanning Tunneling Microscopy (STM) experiments, since it allows 
both topographic imaging and Scanning Tunneling Spectroscopy (STS)
from which information about the local density of states (LDOS)
can be obtained. Interactions stemming from tube-packing or 
tube/substrate/tip can modify the predicted properties of isolated SWNT
and need further study and detailed analysis~\cite{Edu}. 

The study of electron standing-wave (SW) is of fundamental interest as one
addresses directly theoretical and experimental problems connected with
low-dimensional systems (quasi one-dimensional-molecular wires). Aspects as 
Coulomb blockade, localization,  oscillations in the conductivity and the
quantized conductance 
(in units of the conductance quantum $G_0=2e^2/h$=(12.9~kiloohms)$^{-1}$)
of nanotubes have been already observed\cite{Dekker,Heer,devices}.
In this last case, the nanotube conduct current ballistically and 
do not dissipate heat. The transition from
one-dimensional (1D) to zero-dimensional (0D quantum-dot) system can be studied by
looking at different finite-length carbon nanotubes\cite{Venema,Angel}. 
If a conducting nanotube is cut to a finite length, 
the electrons should then display the standing waves characteristic of a 
1D particle-in-a-box model. 
Evidence for 1D quantum confinement was already 
obtained from transport measurements on single-wall 
tubes~\cite{Dekker,Bockrath}, but the standing-wave states 
have been observed only recently in 1D scans of scanning tunneling 
spectroscopy (STS)~\cite{Venema} and described theoretically\cite{Angel}. 

In the present work we present large-scale {\it ab-initio}
and simple tight-binding (TB) calculations
of the electronic properties (density of states and STM imaging)
of different carbon nanotube structures: chiral and non-chiral SWNT,
three-layer MWNT and a three-tube bundle (the simplest nanotube rope),
tubes with defects and finite-length tubes.
The STM-images show no drastic dependence
on the different local environment of the tube 
and can be easily described in terms of a simple TB model. This is
important for routine studies of geometrical effects in the
experimental images\cite{Angel,Lambin}.  In the case of 
finite armchair tubes deposited on a gold surface 
we present the energies and 
three-dimensional shapes of 1D-standing-wave states, that can be
characterized in terms of a simple `catalog' consisting of only four
different patterns.  Furthermore, we
confirm that all semiconducting tubes with similar diameters have a similar 
DOS around the Fermi level\cite{White,Charlier} but with some particular
cases against this rule. The structure in the DOS for the
smaller semiconducting and metallic nanotubes
shows a dependence not only on the nanotube diameter but also on the 
nanotube chirality. 
The interaction between tubes in a bundle tends to open a small pseudogap in 
the Fermi level\cite{Delaney} and to smooth out the peak structure in the DOS,
making more clear the electron-hole asymmetry.

\section{Theoretical model}

Briefly, we have performed the {\it ab-initio} calculations using the 
standard plane-wave pseudopotential total-energy scheme \cite{pwlda,pwlda1} 
in the local density approximation (LDA)~\cite{Cep}
to the exchange correlation potential. {\it{Ab-initio}} norm-conserving 
nonlocal ionic pseudopotentials have been generated by the soft-pseudopotential
method of Troullier and Martins \cite{pseu1}. The LDA wave functions
were expanded in plane-waves up to a 48-Ry cutoff 
(see refs.~\cite{pwlda,pwlda1} for details of the method). 
When studying finite length tubes, 
the large unit cell together with the large number of atoms 
involved ($\simeq$ 1000) makes the plane-wave calculation prohibitive.
 In this case we made calculations in a localized
atomic-orbital basis set~\cite{siesta}.
The scheme includes order-N-algorithms that allows relatively fast 
calculations with several hundred of atoms in the unit cell and has been 
already applied successfully in studying electronic, structural
and STM images of carbon-nanotubes\cite{Angel,fonones}.

The STM-topographic images were simulated within the Tersoff-Hamann 
theory\cite{Tersoff}. 
In this model the tip is not taken into account explicitly,
therefore convolution effects due to the tip shape are neglected.
The STM current for an external applied bias voltage $V$ is proportional to the spatially-resolved
local-density of states (LDOS)
integrated between the Fermi level of the tip and sample
\begin{equation}
I({\bf r},V) = \int_{\epsilon_F-V}^{\epsilon_F} dE~\rho_{LDOS}({\bf r},E) \; ,
\end{equation}
with
\begin{equation}
\rho_{LDOS}({\bf r},E)= \sum_i \mid \psi_i({\bf r}) \mid^2 \delta(E_i-E) \;,
\end{equation}
where $\psi_i$,$E_i$ are the electron wavefunction and eigenvalue of state $i$, respectively. We then approximate the constant current images as isosurfaces of 
$I({\bf r},V.)$.
A description where the tip is treated as a single $s$-atom has been shown to be important in describing the correct shape and relative intensities in the topographic images\cite{Lambin}. Moreover, the differential conductance 
$dI/dV$, as measured in scanning-tunneling-spectroscopy experiments, gives 
direct information of the wave-function square-amplitude and, 
therefore, has direct access to the electronic level structure.
In practice, two limiting factors determine the imaging of
standing-wave patterns: one extrinsic factor due to the
experimental energy resolution (smaller than the molecular
level spacing) and another intrinsic factor related to the electron lifetime
of the molecular-state (SW).
Recent calculations shows that conduction
electrons in armchair nanotubes have very large electron mean free paths 
resulting in exceptional ballistic transport and localization lengths of 
10$\mu$m~\cite{Todorov}. Therefore the main 
scattering at low experimental temperatures (T$\sim$4~K) for
short tubes stems from the tube-boundaries and/or defects
and should be small, making the observation of single molecular orbitals 
possible\cite{Venema,Angel}.

For armchair tubes we have performed a complete study of the STS 
images within the usual tight-binding model for a $\pi$-bonded 
graphene sheet~\cite{book}. In this model we retain only nearest
neighbor interaction between $p_z$-orbitals oriented perpendicular 
to the tube axis. The Hamiltonian is $H_{ij}$=$-\gamma_0$ for nearest 
neighbor atoms, and $H_{ij}$=$0$ otherwise and it is known to provide 
an excellent description of the low energy
features for isolated armchair nanotubes~\cite{Hamada,Mintmire}, 
when  $\gamma_0 \sim$ 2.7~eV.
This model can be solved analytically and in Fig.~1
we present the four wavefunction patterns obtained by combining 
the bonding and antibonding solutions with the sine and cosine envelopes for electronic states close to the Fermi level.
They constitute the ``{\it complete catalog}''
of the STS patterns to be expected
near the Fermi level at the center of the armchair nanotube. 
This simple prediction of the TB model is confirmed by {\it ab-initio}
calculations~\cite{Angel} to be presented below and give us confidence 
in using the TB model to describe low-energy properties of other metallic and 
semiconducting tubes. 
The predicted  `catalog' of only {\it four} 3D STM patterns 
should be confirmed experimentally and extended
to other tube chiralities in the future.

\begin{figure}[h]
\centerline{
\epsfxsize=7.7cm
\epsffile{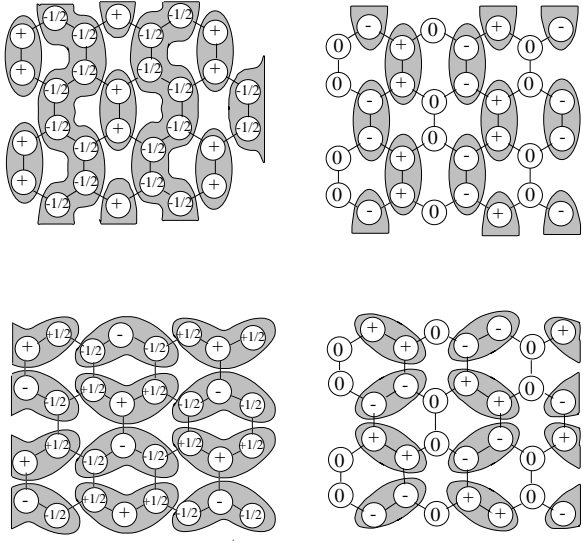}}
\begin{center}
\parbox{10cm}{
\caption[]{\footnotesize{
Schematic tight-binding catalog of
STS images for an armchair nanotube close to the Fermi level. It
correspond to the bonding and antibonding $sin(kz)$ and $cos(kz)$
solutions in a 1D-confinement box model (the wavefunction values are indicated by the + and - symbols).
The general solution for a finite length $L$ tube
is given by a combination of sine and cosine functions with the same
character. In fact a change in length from $L$
to $L\pm a/2$, $a$ being the lattice parameter along the tube axis,
corresponds to a shift in the image pattern. 
The same scheme holds for supported tubes on a gold (111) substrate.
}}}
\end{center}
\end{figure}

\section{Theoretical STM images}

\subsection{Infinite long tubes: SWNT, MWNT and bundles}

Scanning tunneling microscopy  experiments have resolved the atomic 
structure and confirmed the predicted interplay between geometry and 
electronic properties~\cite{STM}.  However, the determination of the 
diameter of the nanotube is not straightforward due to tip-convolution 
effects and operation mode\cite{AFM}.
The chiral angle can be affected by mechanical distortions\cite{Johnson} 
and by the geometry of the STM experiment in obtaining the topographic image:
the cylindrical geometry of the nanotube produces a geometrical distortion 
of  the image stretched in the direction perpendicular to the tube 
axis\cite{Lambin}. Our {\it ab-initio} calculations for the shape
of free-standing isolated SWNTs  support the results 
obtained by a much simpler tight-binding model\cite{Lambin}.
However the present work give some 
information about the influence of tube-tube and tube-substrate interaction 
in bundles of tubes (nanotube ropes) and in multiwall-nanotubes that are beyond 
the capabilities of semi-empirical models. 

\begin{figure}[h]
\centerline{
\epsfxsize=4.5cm
\epsffile{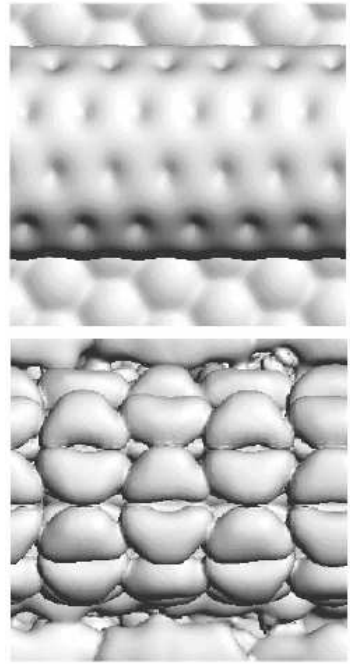}}
\begin{center}
\parbox{10cm}{
\caption[]{\footnotesize{
Ab initio calculation for a (5,5) carbon nanotube supported on Au(111): Top, 
STM-topographic image for an applied voltage of 2 eV.
Bottom, standing-wave pattern of the highest-occupied-molecular-orbital (HOMO). This SW-pattern fits one of the catalog of STS-images in
Fig. 1, showing the marginal role of the substrate in defining the STS-images of
states close to the Fermi level of the supported tube.
}}}
\end{center}
\end{figure}

Many STM experiments are performed on supported tubes on substrates, 
therefore it is important to get insight about the role played by the
substrate in the experimental images. 
We performed first-principles calculations for a (5,5) carbon nanotube 
supported on a Au(111) surface~\cite{Au} as in the experiments\cite{Venema}.
The gold substrate modifies the spectrum in several ways.
(i) It opens a small ``pseudogap" in the tube states at the Fermi
level whenever the symmetries of the tube are
not respected by the gold substrate~\cite{Delaney}, as in this case were the
mirror symmetry of nanotube is destroyed by the substrate.
This symmetry is responsible for the metallic behavior of the 
armchair nanotubes and breaking it would alter its electronic properties.
(ii) It shifts the Fermi level, producing a transfer from the gold to 
the nanotube and a quite strong tube-substrate bonding that prevents the 
tube from moving (binding energy of $\sim$ 1.2 eV per tube unit cell).
More important for STM-imaging is the fact that the charge transfer
does not change the SWNT image catalog of Fig. 1.
This can be seen in Fig. 2 were we show the computed charge density and STS
images for a state close to the Fermi level for
the supported nanotube. Furthermore, by looking in detail at all states
close to the Fermi level, we observe that the shape and general form of 
the wavefunctions 
matched the previously discussed patterns for unsupported tubes\cite{Angel}.
Although these studies were performed for periodic tubes,
we expect the same results to hold for finite-length supported tubes
(see below and ref.~\cite{Angel} for more details). The fact that 
states a few tenths of an eV above/below the Fermi level  are not
resolved experimentally~\cite{Venema} supports  
our observation that the validity of the
two-band-model for the electronic structure of an isolated SWNT
is destroyed due to the strong interaction between 
the unoccupied ascending band of the tube and the Au(111) surface state. 

\begin{figure}[h]
\centerline{
\epsfxsize=7.0cm
\epsffile{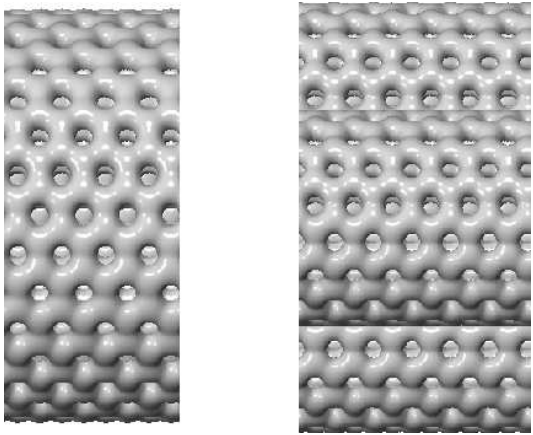}}
\begin{center}
\parbox{10cm}{
\caption[]{\footnotesize{STM-topographic image 
computed at an external applied bias of +0.5~eV.
Left, for a MWNT formed by three concentric shells of armchair 
tubes:  (5,5)+(10,10)+(15,15) of
diameters 0.68, 1.36 and 2.05~nm, respectively.
Right, corresponds to a tube bundle formed by three (8,8) 
armchair tubes (of $\sim$ 1.1~nm diameter).
In all cases the tube-tube distance is very close 
to the corresponding graphitic value (0.34~nm).
}}}
\end{center}
\end{figure}

Now we focus on how tube-tube interactions modify the previous STM patterns.
In Fig.~3 we present the results of the calculated 3D-STM images obtained for
an external voltage of +0.5~eV for a bundle made of three (8,8) nanotubes 
and for a three-wall MWNT made of (5,5), (10,10) and (15,15)
concentric nanotubes. We have only relaxed the structure of the 
isolated SWNTs. Then the
bundle is formed by packing the SWNTs keeping
the tube distance close to the graphitic value.
We checked that changing the polarity of the applied voltage
does not introduce appreciable changes in the STM-topographic image.
Even for the isolated tubes, the curvature makes somehow
visible the asymmetry of the two inequivalent carbon-atoms in a graphitic
structure. This effect is not of great importance for the topmost atoms and
gets more striking as we move along the circumference of the tube.
However, the atomic corrugation would be hidden by the much larger
geometrical corrugation of the object that the tip must follow.
This is relevant in getting the usual 2D-STM-topographic maps~\cite{Lambin}.
We are extending this study to more complex situations including
tip-convolution effects\cite{Meunier}.

\begin{figure}[h]
\centerline{
\epsfxsize=7.0cm
\epsffile{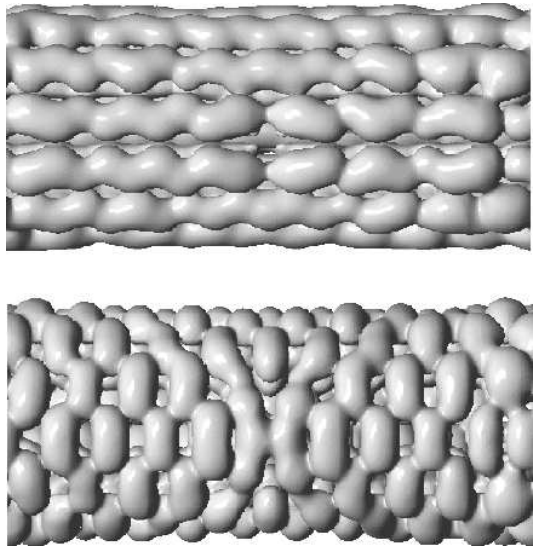}}
\begin{center}
\parbox{10cm}{
\caption[]{\footnotesize{ STM topographic images of a (6,6) carbon nanotube 
with a pentagon/heptagon-pair oriented along the tube circumference with 
the pentagons aligned with the tube axis. Top (bottom)
figure corresponds to an external applied voltage of -0.5 (+0.5)~eV. 
We see a strong dependence on the bias voltage reproducing some of the TB-patterns of the complete catalog shown in Fig.~1. 
}}}
\end{center}
\end{figure}

\subsection{Topological defects and caps}
 
Defects are expected to be present in nanotubes in several forms: doping
and topological defects, as well as hybridisation and incomplete bonding.
The presence of such defects could alter substantially both the electronic 
and elastic properties of nanotubes. Topological defects as the Stone-Wales 
transformation (pentagon/heptagon pair obtained by a $\pi/2$ rotation of one
bond in a four-hexagon complex) are difficult to be observed in transmission
electron microscope experiments, given that they conserve the curvature of 
the tube. Depending on the tube symmetry, either metalization or band gap 
opening can result. In the case of (n,n) tubes a line of allowed 
{\em k\/}-values runs from $\gamma_0$
to $K$ in the graphite extended zone, moving the Fermi level along this
line from $K$ to $\gamma_0$ upon introduction of rotated-bond defects 
should not open a gap. Instead, an increase of approximately 25~\%\ in the 
density of states at the Fermi level should be expected, as the 
$\pi-\pi^*$ band dispersion decreases\cite{Vin}. Therefore, STM experiments
might give valuable information about these topological effects.
Along this line we present in Fig.~4 the computed image of a (6,6) nanotube 
with a topological Stone-Wales defect for two different applied 
voltages $\pm$0.5~eV. The geometry of the defect has been 
relaxed by assuming periodic boundary conditions  with a unit-cell length 
of 2.5~nm. The total-charge density has an accumulation (deficiency) of
charge on the pentagon (heptagon) site that has importance in the use
of nanotubes for electro-chemical applications~\cite{reactivity}.

Some clear effects are observed from the calculations shown in Fig.~4:
(i) the STS-images fall onto one of the patterns predicted by the 
simple TB-model in Fig.~1 and different external applied voltages selects
specific states of the catalog. (ii) Due to the mirror symmetry along a plane 
perpendicular to the tube axis of the pentagon/heptagon-pair defect, 
the wavefunction pattern is identical to both sides of the defect. This would
not be the case for other relative orientations of the defect.
A more complex structure has been predicted in ref.\cite{Lambin} for a single 
pentagon/heptagon defect joining a metallic and semiconducting tube; 
the perturbation dies out much faster on the semiconducting side than 
on the metallic one. We note that localized-defect states as well as
tip-states have already been observed in STS experiments\cite{tip} and 
our calculations show that the Stone-Wales defect can be, 
indeed, experimentally accessible by STS measurements and 3D-mapping.
Physically we can interpret these patterns as the result of an interference 
effect between the scattered electron-waves by the defect. 
The STM pattern would be different depending on the symmetry of the defect.

\begin{figure}[h]
\centerline{
\epsfxsize=8.0cm
\epsffile{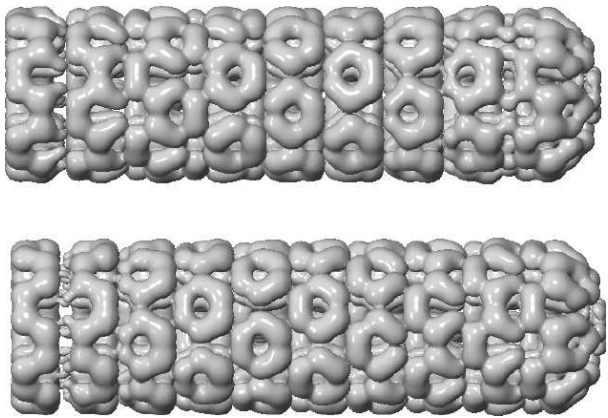}}
\begin{center}
\parbox{10cm}{
\caption[]{\footnotesize{ STM-topographic image  for a (6,6) capped carbon 
nanotube of $\sim$3.5~nm length. The cap consist of a perfect half-fullerene 
on one end and open-end saturated with H. 
Top (bottom) figure corresponds to an external applied voltage of -0.5 (+0.5)~eV
that in the present case corresponds  basically to sample the LUMO (HOMO) 
states of the finite system. 
}}}
\end{center}
\end{figure}

To conclude this section, we show in Fig.~5 the computed STM-topographic
images for a finite-capped (6,6) tube. The capped geometry consist of a 
perfect half-fullerene (C$_{60}$) on one end and on the other and open-end 
saturated with hydrogen atoms to avoid the formation of dangling bond-states.
Again we see the formation of electron standing-wave formed by a constructive 
interference between the electronic states around the Fermi level and its
reflection on the tube boundaries. The obtained pattern can be 
described as a linear combination of the simplest TB-patterns with 
some chiral-pattern that depends on the geometrical orientation of 
the pentagons in the cap. We note again a clear dependence of 
the topographic-image on the external applied voltage due to the different
electronic state observed. The scattering at the tube-boundaries produces
the formation of different standing-wave patterns with a rotational
symmetry dictated by the relative position of the pentagons in the cap with
respect to the hexagonal network.
Standing-wave patterns are also present in the case
of finite-length tubes (see below).
These complex patterns are similar to the ones expected to 
be observed for chiral carbon nanotubes.

\subsection{Finite size effects: electron standing-waves}

More detailed information about the electronic structure of 1D-quantum wires can be directly obtained in STS experiments by mapping the 
1D-confinement of electrons in the nanotube structure. This can be achieved
by cutting the tube to a finite-length~\cite{cut} what reduces the
periodic band-structure to a discrete set of molecular levels\cite{Klein}
that can now be imaged by STM~\cite{Venema,corrals}. In this simple  
scenario of a 1D particle-in-a-box model, 
a tube of length $L$  has a  set of allowed k's given by 
$k=n\pi/L$ ($n$ integer). Taking the Fermi level of the tube at the
single graphene-sheet value of $k_F=\frac{2\pi}{3a}$,
the wavefunctions close to the Fermi level will exhibit a 
periodic pattern with a wavelength of $\lambda_F=3a=0.74$~nm, as
observed in STS measurements~\cite{Venema}. Although, this basic standing wave 
observation can be explained in terms of the simple 1D particle-in-a-box model,
further insight is needed to understand their
energy and three-dimensional shape.

The four-image pattern presented in Fig. 1 is fully confirmed by 
extensive ab-initio calculations for SWNT
of different lengths and diameters 
isolated and supported on gold Au(111)~\cite{Angel,Au}.
In spite of the `weak' disorder
introduced by the interaction with the substrate,
our results indicate
that the electron-images close to the Fermi level fits the same pattern as
that of isolated tubes. Furthermore,
the experimentally observed effect of peak pairing~\cite{Venema}
is explained by the asymmetric shapes of the lobes in the catalog 
of Fig. \cite{Angel}. An example of the STM-images obtained for a supported
(5,5) tube on gold was
presented in Fig.~2 where the topographic image for an applied voltage of 1eV (top) and STS-image of the HOMO-state (bottom) are given.
We confirm the observation of 
a standing-wave modulation of 0.74 nm near the Fermi level. 
Also, the peak pairing observed in the experimental scans~\cite{Venema} 
is understood in sight of our calculations~\cite{Angel}.

We find that the electronic-state 
images can be understood in terms of the simple TB-model, 
which offers a catalog of just four image patterns~\cite{Angel}. However, 
the associated energies are very sensitive to different effects beyond that 
model: the relaxed geometry, the electronic self-consistency 
in the finite tubes, and the interaction with the substrate
In general, the value of the HOMO-LUMO gap decreases with increasing tube length not monotonically but exhibiting a well 
defined oscillation that is related to the bonding character of the
HOMO and LUMO orbitals~\cite{Klein}. By increasing the 
tube-length we observe a smooth transition from an energy level structure 
characteristic of a molecular-wire (zero-dimensional system)
to that of a delocalized one-dimensional system, that seems to be
complete for tube-lengths of the order or larger than 5~nm~\cite{Klein}.
We conclude that tube curvature, termination, structural 
relaxation, and substrate interaction {\em do not} alter significantly
the simple TB-STM patterns, which therefore should
indeed be observed experimentally\cite{Angel}.
However, this is not the case for the
level structure that is very relevant for the transport properties
of the recently proposed electronic devices based on supported nanotubes.

A general remark should now be made. We have presented all throughout this work 
3D-STM-images corresponding to a constant current operational mode in the STM.
This data can be accessible experimentally,
however the usual published data correspond to 2D-topographic maps of the
nanotubes. When bringing our data to this 2D-maps care has to be taken due
to a geometrical distortion induced by the tip-tube geometry~\cite{Lambin}.
These geometrical induced distortions prompt us to present our computed 
STM-images in a 3D-format.

\section{Density-of-states: STS-spectroscopy}

A better knowledge of the electronic properties of carbon nanotubes
can be obtained by complementing the previous 
STM-images with spectroscopic data, i.e, density-of-states (DOS) calculations.
The DOS can give a direct information about the metallic/semiconducting
behavior of the tubes as well as particular insight into the tube-tube or 
tube-substrate interactions. Information about structural properties and 
local environment for a carbon or composite nanotube can be 
extracted from the computed LDOS\cite{Edu}.
In a recent work, the connection between 
tube-diameter and low-energy features in the DOS 
has been pointed out\cite{White}. The fact that the electronic DOS for each
metallic or semiconducting tube is
practically independent of the nanotube chirality, is in qualitative 
agreement with STS experiments~\cite{STM}. The simple $\pi$-electron TB model
was used to get this general correlation between tube-diameter and
features in the DOS~\cite{White}. As we show below, the neglect of 
curvature effects and $\sigma$-$\pi$ hybridisation leads to quantitative changes
in the DOS in both peak energies and intensities. Therefore, the TB-results
are only valid for states a few tenths of an eV above or below the 
Fermi level, and ab-initio calculations are needed to address the validity 
of this simple model.

\begin{figure}[h]\label{fig4}
\centerline{
\epsfxsize=7.5cm
\epsffile{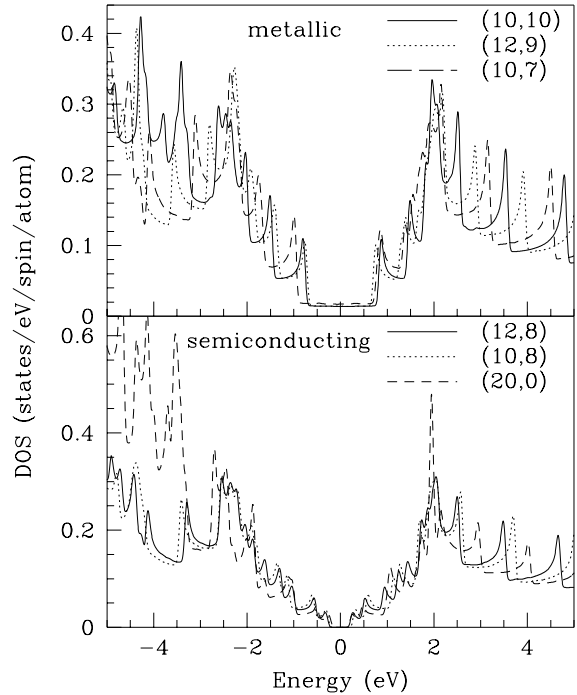}}
\begin{center}
\parbox{10cm}{
\caption[]{\footnotesize{Ab initio DOS for different metallic and 
semiconducting tubes
with diameters in the range of 1-2~nm, namely: 1.17, 1.36, 1.44, 1.23, 1.37 and
1.58~nm for the (10,7), (10,10), (12,9), (10,8), (12,8) and (20,0) nanotubes,
respectively. Spikes in the DOS stems from the van Hove singularities of the
nanotube 1D-band-structure and they play an important role in the
description of resonant-Raman scattering experiments. All DOS are normalized to the number of atoms in the nanotube unit cell.
}}}
\end{center}
\end{figure}

In an experimental characterization of a carbon nanotube, the DOS plays a key role as can be mapped over a wide range of applied bias.  For this reason we plot in Fig.~6 the computed ab-initio DOS  for a set of chiral 
and non-chiral tubes with diameters around the experimental value of 1.3 nm
(between 1.1 and 1.6 nm). The following conclusions can be 
extracted from the figure:
(i) in metallic tubes the
plateau around the Fermi level depends on both tube diameter and, to a lesser
extent, on tube-chirality. For almost all
tubules with $\sim$1.3~nm diameter the metallic-plateau is about 1.7-2.0~eV.
This data is of importance is discriminating metallic and semiconducting tubes
in resonant-Raman scattering experiments~\cite{Rao,Pimenta}. The non-armchair
tubes belonging to the metallic group are indeed quasimetallic with a
extremely small gap introduced at the Fermi level by curvature effects. 
(ii) The electron-hole symmetry of the TB-model is no longer valid even for the first spikes in the
DOS (see the clear example of the (10,7) metallic tube).
This effect gets more clear as the nanotube radius is reduced or/and as
we move away from the Fermi level. The separation between van Hove 
singularities is also slightly different for both conduction and valence states. 
(iii) The direct connection between the diameter and the structure of the spikes
in the DOS is not always clear. Note for example that the semiconducting
(12,8) and (20,0) nanotubes have very similar DOS close to the Fermi level. 
However their diameters are 1.37 and 1.58 nm, respectively. The same holds
for the metallic (12,9) and (10,10) tubes with very similar 
``metallic-plateau", but diameters of 1.44 and 1.36~nm, respectively.
Then, although the proposal in ref.~\cite{White} is very appealing,  its
practical application to discern the tube-diameter is doubtful in its
spatial resolution (not better than 0.15~nm for the diameter). In conclusion, to
elucidate the geometrical structure of a carbon nanotube we would have
to combine spectroscopic and imaging techniques in STM experiments
together with simulations. 

\begin{figure}[h]\label{fig5}
\centerline{
\epsfxsize=8.0cm
\epsffile{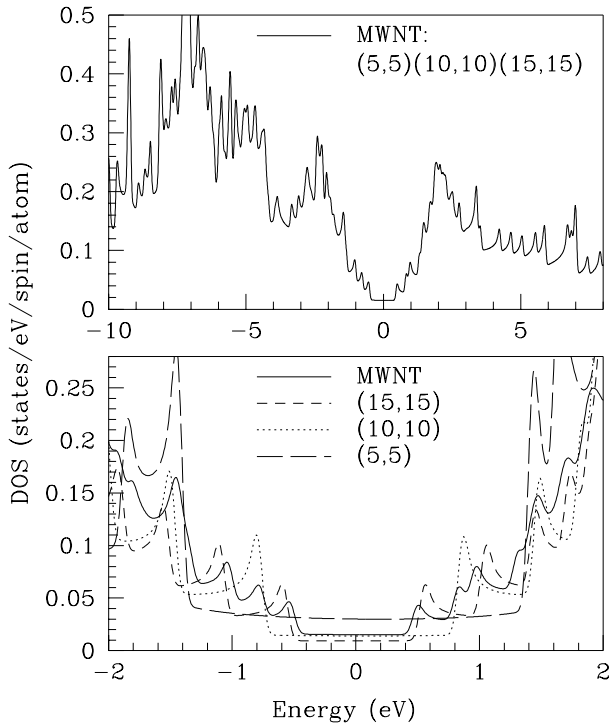}}
\begin{center}
\parbox{10cm}{
\caption[]{\footnotesize{DOS for a MWNT formed by three concentric armchair
tubes: (5,5)+(10,10)+(15,15). Bottom, comparison of the DOS for the MWNT with
each of the component nanotube DOS magnified around the Fermi level. Each DOS is normalized to the number of atoms in the unit cell. 
}}}
\end{center}
\end{figure}

To get insight into tube-tube interaction in MWNT we plot in Fig.~7 the
DOS for a MWNT formed by three concentric armchair tubes such that 
the inter-tube distance is close to the graphitic value. 
We see that the low-energy structure seems to give
information about the number of layers in the tube, 
however this identification gets more complicated when non-commensurate 
metallic or semiconducting tubes participate as main building blocks of 
the MWNT. We obtain the expected result that the metallic-plateau of the MWNT is 
mainly controlled by the outer tube (similarly to the 
STM-images in Fig.~3). The interaction among tubes being weak, 
only shifts a little bit the
position of the van Hove singularities in the MWNT with respect to the SWNT. 
This shift is larger for the conduction states making the electron-hole
asymmetry more clear. Modifications due to a finite number of
graphene layers with different stacking appear in the DOS as 
peak-structure in the 1-4 eV range. This study was done on
a set of graphene layers with different stacking sequence and interlayer 
distances. Above 3~eV we see the appearance of interlayer and
surface states.
The results will be reported elsewhere\cite{Edu,Carrol}.

\begin{figure}[h]\label{fig6}
\centerline{
\epsfxsize=8.0cm
\epsffile{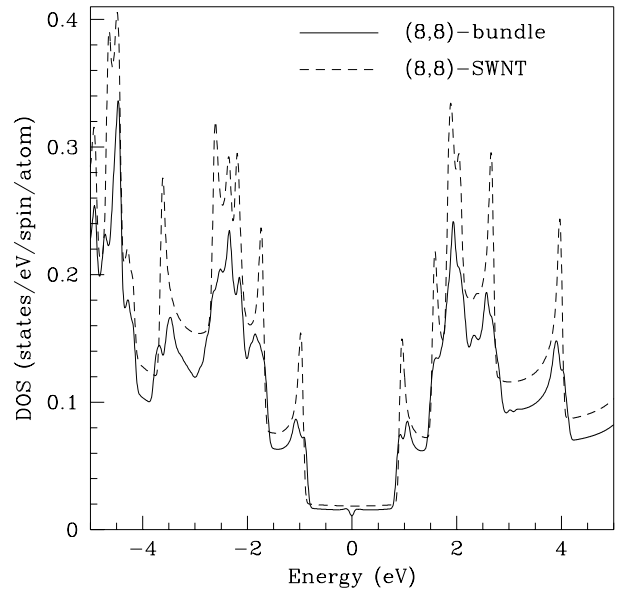}}
\begin{center}
\parbox{10cm}{
\caption[]{\footnotesize{DOS for a small nanotube-rope (bundle)
formed by three (8,8) SWNT (1.09~nm diameter)
 packed in a triangular lattice with an intertube distance 
of 0.345~nm. We clearly see the opening of a ``pseudogap" of about $\sim$0.1~eV
around the Fermi level. We compare the results for the bundle with the 
DOS for an isolated (8,8) SWNT (dashed line).
}}}
\end{center}
\end{figure}

In Fig.~8 calculations for a bundle or nanotube-rope constituted by 
three (8,8) SWNT packed on an equilateral triangle network with 0.345~nm 
intertube distance, aimed to 
better understand the role of tube-tube interactions in the DOS.
In this case the intertube 
interaction clearly modifies the spectra seen in the 
DOS. (i) It opens a ``pseudogap" close to the Fermi level as already
predicted for random oriented nanotube ropes\cite{Delaney} 
(pseudogap of $\sim$~0.1 eV). The bundle remains metallic. 
(ii) It makes the electron-hole asymmetry in the DOS more accentuated 
and the spike structure of the van Hove singularities is smooth out. 
The fact that the position in energy of the peaks is not strongly modify 
explains the success of using isolated SWNT spectra to describe the experimental
data~\cite{STM}. However the shape of the spectra (relative intensities)
is strongly affected by tube-tube interactions as it is clearly seen in Fig.~8. 

To end this section it is worth to discuss these results in terms of the simple
$\pi$-electron TB model.
Within this simple model the DOS in the vicinity of the Fermi level is
symmetric (electron-hole symmetry) and can be
expressed in terms of a universal function that depends only in whether the tube
is metallic or semiconducting~\cite{White}. In terms of the nearest neighbor
overlap energy $\gamma_0$ we have that, for a semiconducting tube, the band-gap
is given by 
\begin{equation}
E_g = \frac{2\gamma_0 a_{C-C}}{D}  \;,
\end{equation}
where $a_{C-C}$ is the carbon-carbon
bond-length ($\sim$~1.42\AA) and $D$ is the nanotube diameter. 
In the case of metallic tubes, the metallic plateau ($E_{met}$),
given by distance between the two van Hove singularities above and
below the Fermi level, is 
\begin{equation}
E_{met} = \frac{6\gamma_0 a_{C-C}}{D} \;.
\end{equation}
In both cases the distance between consecutive conduction or valence van Hove
singularities is given by $\Delta E= \frac{3\gamma_0 a_{C-C}}{D}$.
This $\gamma_0$ parameter plays an important
role in the experimental analysis of their electronic structure data. In fact,
a fit to STS experiments~\cite{STM} give a value of $\gamma_0$=2.7~eV, whereas
the fit to resonant Raman scattering experiments on metallic carbon 
nanotubes~\cite{Pimenta} gives $\gamma_0$=2.95$\pm$0.05~eV. This indirect
estimation is in quite good agreement with the direct measurement by STS, and
both are smaller than the $\gamma_0$=3.16~eV value for graphite~\cite{book1}. 
These results stem from the fact that in the TB model we have a radial 
dependence of the bandstructure with
respect to the Fermi level ($K$-point in the Brillouin Zone of the graphene
sheet). Our results summarized in Figs.~6,7 and 8 show that the value of 
$\gamma_0$ is not unique due to the anisotropy of the DOS in both peak positions
and intensities. The small band-gap of semiconducting tubes makes them as the
best candidates to experimentally determine the value of $\gamma_0$ from eqn.~(3).
The computed values go from 2.77 to 2.95~eV for tubes with diameters of 
1.23 and 1.58~nm, respectively. Smaller values are obtained for metallic tubes
when fitting the metallic plateau to eqn.~(4) (from 2.32 to 2.75~eV for 
diameters of 0.68 to 2.04~nm, respectively). In general, the value of $\gamma_0$
increases with increasing nanotube diameter. 
Furthermore, the interaction between tubes also modify this parameter by as much
as 10\%~\cite{comment_bundle}.
In fact, the electron-hole asymmetry in the DOS is a measure of the 
curvature effects, intertube interactions and anisotropy in the band-structure.
Note that due to the fact that the 
conduction  wave functions are spatially more extended than the valence ones,
the unoccupied part of the DOS is more sensitive to
any external perturbation (this is clearly seen in Figs.~6,7 and 8)

\section{Summary}

In conclusion, we have presented a study of the STM images and electronic 
properties of carbon nanotubes in different environments. The present
technique opens a new way to address the analysis of the interplay between 
electronic and geometric properties in carbon nanotubes. The method has been 
applied to SWNT, MWNT and nanotube-ropes as well as  
finite-length armchair nanotubes. 
Within a very simple TB-model we
have characterized the complete catalog  of STM-images as 
consisting of four images. These results are further confirmed by ab-initio
calculations.
The gold substrate does not alter the main images of the isolated tubes even if the tubes are quite strongly bound to the substrate by charge
transfer. The computed images are in very good agreement with 
experiments in both wavelength of the standing
wave (0.75 nm) as well as in inner details (pairing).
Surface states related to the boundary of the
tubes are observed to appear within 1~eV above the Fermi level. These states
could play a role in the field-emission process and chemical activity
and for the description of the electronic structure of other chiral tubes
(semiconducting) where they can be very close to the bottom of the conduction
band. Moreover, different STS-patterns are expected for chiral tubes that can   be used as a further check of the `particle-in-a-box' model.  
Spectroscopic studies based on looking at the DOS gives additional information
about the metal/semiconductor character, packing and tube-tube interactions. 
In this respect, most of the peak structures for the composite-DOS can be
understood in terms of the isolated SWNTs DOS. However, subtle but important
differences appear in the intensity and shape of the peaks as well as in the
opening of a pseudogap close to the Fermi level and modification of the effective
metallic plateau.
These studies nicely complement the STM-imaging and provide us with a lot 
of information that can be checked by experiments.

\vspace*{0.5 cm}

{\em Acknowledgments: }
We acknowledge financial support from DGES (Grants: PB95-0720 and PB95-0202)
European Community TMR contract ERBFMRX-CT96-0067 (DG12-MIHT).
Computer time was provided by the 
C$^4$ (Centre de Computaci\'o i Comunicacions de Catalunya).
We thank C. Dekker, L.C. Venema and D.L. Carrol for sharing their 
experiments with us prior to publication and for enlightening
discussions. We also benefited from fruitful collaborations 
with E. Artacho, E. Hern\'andez, 
Ph. Lambin, M.J. L\'opez, V. Meunier, P. Ordej\'on, 
D. S\'anchez-Portal and J.M. Soler.


\begin{references}
\bibitem{Au} The Au(111) substrate is described by a six layers slab.
During relaxation the near-neighbor atomic distance
of the gold bottom layer was fixed to the bulk value (i.e, 2.88 \AA).
The computed work function of the surface is 
5.01 eV close to the experimental work function $\sim$ 5.3~eV.
On top of the relaxed surface we deposited
a relaxed (5,5) nanotube with its axis pointing 
along the (1\-10) surface direction.  Although there is no
experimental evidence of a special orientation of the supported tubes, 
we chose this direction only for practical purposes in order to
get a commensurate tube-substrate unit cell. 
The tube-substrate distance that minimizes the
total energy is $\sim$~2.5 \AA.
\bibitem{comment_bundle} The determination of $\gamma_0$ turns out to depend on
the experimental conditions and model used to extract its value (it is not
longer unique but rather model-dependent). Note that the DOS for the
nanotube bundle in Fig.~8  shows a complex spike structure
above and below the Fermi level. Depending on which peak-pair is used to fit
eqn.~(4) we get variations of 10\% above and below the $\gamma_0$ value for the
isolated (8,8) SWNT ($\gamma_0 \simeq$2.5~eV).
\end{references}
\end{document}